\documentclass[times,authoryear]{elsarticle}

\usepackage{jasr}
\usepackage{framed,multirow}
\usepackage{subcaption}
\usepackage{amssymb}
\usepackage{latexsym}
\usepackage[T1]{fontenc}

\usepackage{url}
\usepackage{xcolor}
\definecolor{newcolor}{rgb}{.8,.349,.1}

\usepackage[citebordercolor=white]{hyperref}
\usepackage{amsthm,amsmath}

\journal{Advances in Space Research}
 
\begin{document}

\verso{Taran \textit{etal}}

\begin{frontmatter}

\title{Effect of  Geomagnetic Storms on a Power Network at Mid Latitudes}%

\author[1]{Somayeh Taran}
\ead{s.taran@znu.ac.ir} 
\author[1]{Nasibe Alipour \corref{cor1}}
\cortext[cor1]{Corresponding author: Nasibe Alipour}
\ead{alipourrad@znu.ac.ir} 
\author[1]{Kourosh Rokni}
\author[2]{S.Hadi Hosseini}
\author[3]{Omid Shekoofa}
\author[1,4]{Hossein Safari \corref{cor2}}
\cortext[cor2]{Corresponding author: Hossein Safari}
\ead{safari@znu.ac.ir}


\address[1]{Department of Physics, Faculty of Science, University of Zanjan, University Blvd., Zanjan, Postal Code: 45371-38791, Zanjan, Iran.}
\address[2]{Department of Electrical Engineering, Faculty of Engineering, University of Zanjan, University Blvd., Zanjan, Postal Code: 45371-38791, Zanjan, Iran.}
\address[3]{Satellite Research Institute, Postal Code: 1997994313, Tehran, Iran}
\address[4]{Observatory, Faculty of Science, University of Zanjan, University Blvd., Zanjan, Postal Code: 45371-38791, Zanjan, Iran.}
\received{--}
\finalform{--}
\accepted{--}
\availableonline{--}
\communicated{--}

\begin{abstract}
Solar activities may disturb the geomagnetic field and impact the power grid via geomagnetically induced currents. We study active and reactive powers as well as the power factor of Iran's power grid transformers (230 kV and 400 kV)  and their correlations with geomagnetic disturbances indices (SYM-H $<$ -30 nT and sizable horizontal geomagnetic field fluctuation) from 19 March 2018 to 20 March 2020. Out of 128,627 cases with a transformer power factor of less than 0.7, we observe that 12,112 samples correlated with SYM-H. Our investigation shows that about 4 percent of two years, the SYM-H has values less than -30 nT. Analysis of high-performance transformers (a power factor greater than 0.7 at 95 percent of working time) shows at least a 55 percent correlation of power factor less than 0.7 and SYM-H less than -30 nT. We observe that the transformers' power factor of Rafsanjan-Kerman and Sarcheshmeh-Kerman (wye connection on 230 kV side) substations  decreased to less than 0.7 and correlated with SYM-H. We show that the reactive power of the Sefidrood-Guilan and Shahid Beheshti-Guilan transformers (wye configurations) increased considerably on 9 January 2020 and positively correlated with SYM-H may produce large GICs at this part of the grid. We observe that the increase in reactive power at the Shahid Beheshti-Guilan substation correlated with the sizable changes in the horizontal field recorded by the Jaipur station. However, more details (temperature and current of transformers) need records to estimate the impact of geomagnetic induction current on the power grid.
\end{abstract}


\end{frontmatter}


\section*{Introduction}
Geomagnetic disturbances (GMDs) appear during interactions of the solar wind plasma or a coronal mass ejection (CME) with the Earth's magnetic field. \citep{jonas2015recent,rozanov2016foreword, taran2019kappa,mate2019carrington,abda2020review, strasburg2020risk,shaikh2022impact}.
GMDs are the time-varying Earth magnetic fields that may generate an electric field via Faraday's law of electromagnetic induction \citep {pirjola1991geomagnetic,pirjola2007space,marshall2011preliminary}.
The geomagnetically induced current (GIC) from an induced electric field on the Earth may affect the technological equipment (such as power grid, telecommunication, and oil pipelines) and create problems in the standard course of their operation \citep{pirjola2000geomagnetically, hejda2005geomagnetically, Ngwira2008,oliveira2017geomagnetically, abda2020review}.

Solar activity shows a main period of eleven-year and a variety of energetic phenomena (such as CME, flares, and flares associated with CMEs) appear in the solar atmosphere. GMDs are more likely correlated with high-energy solar phenomena  \citep{gruzdev2007effect,aschwanden2011self,prokoph2012influence}. Historical evidence shows that large solar storms have also occurred during the minimum activity. The lowest recorded solar activity belongs to the Maunder minimum (1645 - 1715) \citep{eddy1976maunder,usoskin2015maunder}, at which time aurora as evidence of a geomagnetic storm occurrence was recorded at low latitudes such as Japan and China \citep{isobe2019intense}. Therefore, the event of a geomagnetic storm may occur every time, even from maximum to minimum solar activity.

The GIC flow in the power grid is a complex process and more likely depends on the network configuration than the geographical latitudes \citep{Zheng2013}.
\citet{trivedi2007geomagnetically} continuously measured the horizontal component of geomagnetic field changes and by installing a fluxgate directly under the power transmission lines (TL) during a geomagnetic storm in November 2004, recorded a GIC of about 15 A in the Brazilian power grid case study.

\citet{watari2015estimation} used the geomagnetic and geoelectric fields to obtain an empirical equation for GIC estimating in past intense geomagnetic storms in Japan. Although their results indicated that low latitudes are not more affected by GIC, they emphasized that the decline or acceptance of this result needs to study for longer. 
\citet{ebihara2021prediction} used geoelectric disturbances data that occurred in India in 1859 to examine the condition of Japan's power grid in confronting a storm similar to a Carrington-class storm. Applying the convolution theory they showed that the lowest amount of GIC that could flow in such a situation in the Japanese power grid was an impressive amount of about 89$\pm$30 A. The shared latitude of Iran and Japan confirms that the GIC flowing in Iran's power grid is not far from expected.
\cite{gil2021evaluating} used electrical grid failure logs, geomagnetic storm index, and solar activity data to study the geoelectric field variations in Poland for 2010–2014. They showed that severe geomagnetic storms, which may have a solar origin, increase the number of transmission line failures.

The disturbance storm time (\textit{Dst}) index is one of the most widely used features for classifying geomagnetic storms that provides a measure of the intensity of the Earth's magnetic field decreases caused by the ring current. Using the \textit{Dst} index, the GMD storms are divided into weak (-50 nT $<$ \textit{Dst} $<$ -30 nT), moderate (-100 nT $<$ \textit{Dst} $<$ -50 nT), intense (-200 nT $<$ \textit{Dst} $<$ -100 nT), and very intense (\textit{Dst} $<$ -200 nT) storms \citep{nikolaeva2011dependence,valdes2021effects}.
SYM-H is a similar index designed to measure storm intensity and has a time resolution of one minute compared to \textit{Dst} with a one-hour resolution \citep{wanliss2006high}.

Different magnetometer stations calculate SYM-H as the symmetric part of the horizontal component of the magnetic field near the equator \citep{wanliss2006high}. \citet{iyemori2010mid} proposed the SYM-H is a suitable geomagnetic index for mid-latitudes. The strength of the local magnetic field is a practical parameter in studying geomagnetic disturbances. Significant sudden changes in the magnitude of the magnetic field are evidence of a major geomagnetic disturbance event in an area. The SuperMag service provides magnetic field data in local geomagnetic coordinate systems (north, east, down) \citep{gjerloev2012supermag}. \citet{rogers2020global} used SuperMag data to provide a global model for the extreme fluctuations of the geomagnetic field.  

Power grid transformers are one of the most impressive power grid equipment that GICs flow through their neutral point and may cause some damage in them \citep{molinski2002utilities, zhu2014blocking,etemadi2014optimal}.
When the GIC passes through the transformer as a quasi-direct current (0.0001 Hz to 0.1 Hz), it causes half-wave saturation of the transformer core and increases its magnetizing current \citep{girgis2012effects,zhang2014study}. Due to GIC in the network, the voltage may be distorted, and reactive power may increase in transformers \citep{ chen2019power,rajput2021insight}. The PSASP (power system analysis synthesis program) simulation revealed a linear relationship between the reactive power losses of the transformer and the GIC \citep{berge2011impact,kang2019analysis}. Investigating the GIC on 400 kV transformers in Finland showed no complete correlation between the various substations' GIC flows. So estimating the significant GICs consequences on the Finnish power system is complicated \citep{lahtinen2002gic}.

\cite{shetye2013geomagnetic} studied the impact of GMD on American Electric Power and determined the transformers with a high probability of effectiveness under different electric fields. They used the PowerWorld Simulator to calculate the reactive power increases related to the GICs. The lowest electric field estimated for transformer voltage collapse was 13 V/km. Similar studies on the power grid of the North American Eastern Interconnect have also confirmed the increase of reactive power in the presence of GIC flow \citep{overbye2013power}. A study on Siberia's power grid provided a graphical visualization of the electric grid's vulnerability to severe geomagnetic disturbances. Based on the topology of the power grid, voltage level, length of transmission lines, and ground conductivity, \cite{sokolova2018algorithm} presented an algorithm that estimates the GIC flow in the network.

Evaluation of geomagnetic induction current on 275 kV power transformers in Malaysia also showed that GICs flowing in the power system leads to half-cycle saturation of the transformer that may result in reactive power losses and voltage instability of the power grid. Simulation of transformers' active and reactive power response to GIC flow showed that the magnitude of active power decreases and the reactive power increases \citep{zawawi2020evaluation}. This study confirmed the GIC event is not limited to high and mid-latitudes and extends to power systems in low-latitude regions.

\citet{albert2022analysis} analyzed the long-term GIC measurements in transformers for five years in Austria's power grid. Using frequency analysis, they found artificial sources of low-frequency currents and frequent geomagnetic activities in different places. The maximum current measured during the five years was 13.83 A, which occurred on 12 May 2021, concurrent with a geomagnetic storm with a minimum \textit{Dst} of -61 nT. The correlation between the magnetic field and transformer neutral point currents variation during high geomagnetic activity showed that GIC flow is sensitive to specific locations and directions of magnetic field change.

Here, we investigate Iran's transformers' active and reactive powers (230 and 400 kV) and correlations with the geomagnetic fields. We use the geomagnetic fields such as the SYM-H index, and SuperMag magnetic field data recorded close to Iran. We study the effect of geomagnetic storms on Iran's power grid as a sample of a mid-latitude country.

\section*{Effects of GIC on transformers}
\begin{figure}[h]
    \centering
   \includegraphics[width=0.85\textwidth]{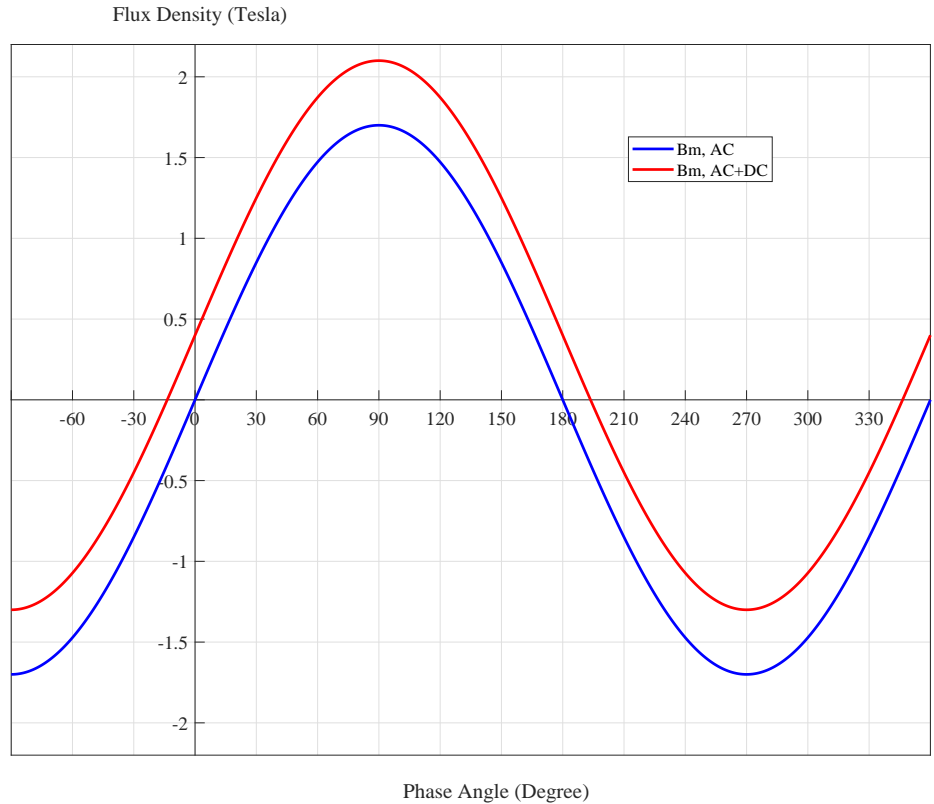}
    \caption{Effect of direct current (DC) on the magnetic flux of a transformer core: Alternative current (AC) Flux density may change in the core with the superposition of the DC flux \citep[see also][]{girgis2012effects}. }
    \label{fig1}
\end{figure}

For a transformer with a wye connection and a grounded neutral point, the GIC can flow in the windings of the transformer. So, we focus on analyzing transformers' connections together and with the ground to identify the possibility  of causing GIC in the network. The magnetic flux flows through the transformer core in one specific direction by exposing a DC on a power transformer. The magnetic flux density for a transformer depends on several factors such as the DC value, the number of turns per windings, and the reluctance of the path of the DC flux \citep{kohli2018impact}.
Due to the presence of DC magnetic flux in a transformer (Figure \ref{fig1}), the AC flux increases from the expected value that is designed to be less than the knee point of the B-H curve of core (nominal) value in the one-half cycle and decreases in the next half cycle \citep{sharma2018impact}.
A sufficiently large DC or high flux density may cause the core flux increases in the magnetic saturation domain in half of one cycle.

\begin{figure}[h]
    \centering
    \includegraphics[width=0.85\textwidth]{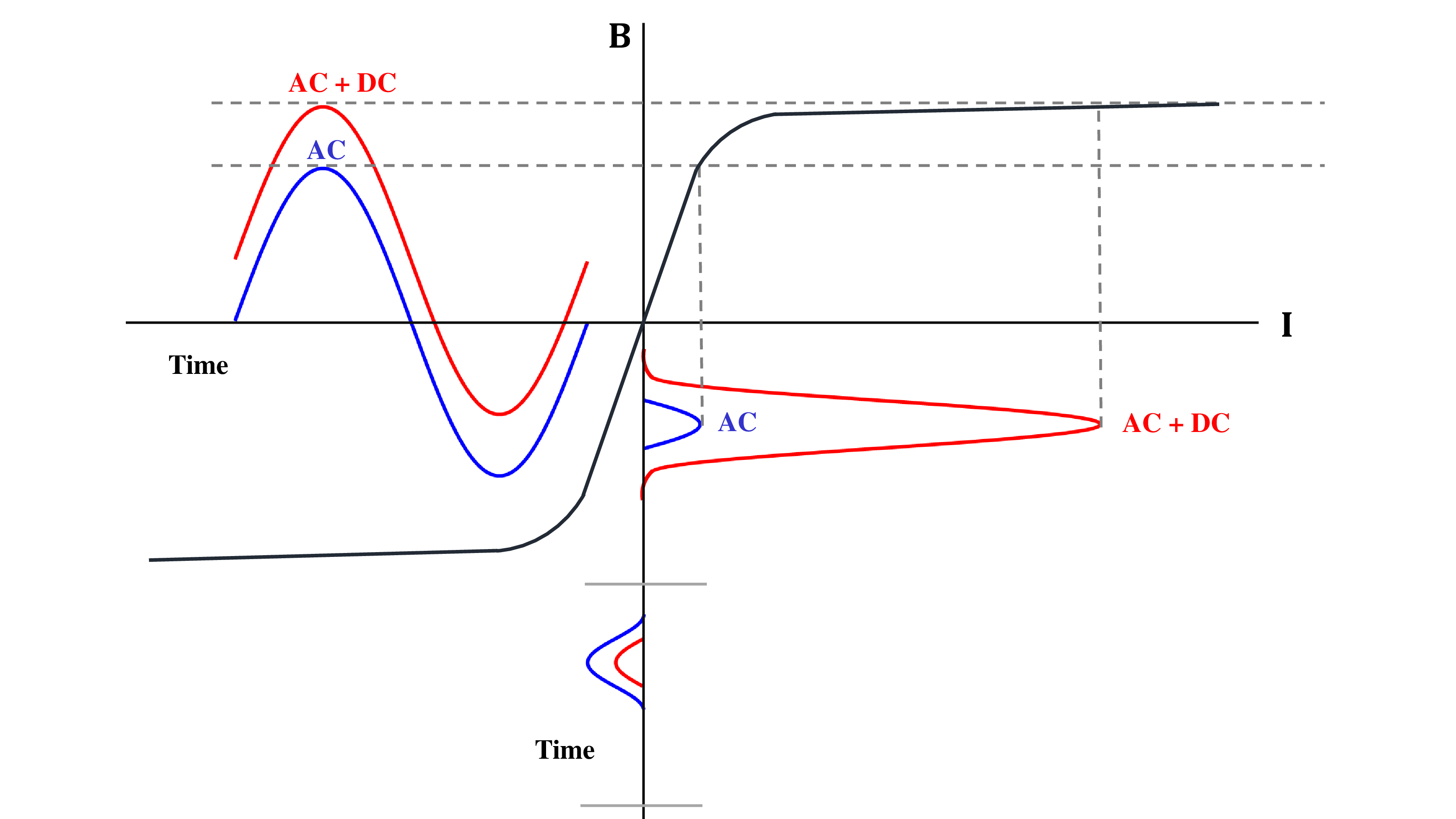}
    \caption{Effect of direct current (DC) on the magnetizing current of a single phase transformer of a transformer core: Semi-saturation cycle of transformer cores \citep[courtesy by][]{girgis2012effects}.}
    \label{fig2}
\end{figure}

Figure \ref{fig2} shows the magnetic flux density ($B$) variation in terms of the magnetization current ($I$) for the core material of a transformer. As shown in the figure, we observe a nonlinear behavior for $I$ as a function of $B$.  
The nonlinear characteristic of the transformer core material causes a magnetization current increase as the transformer core approaches saturation. The transformer is designed to operate below saturation under standard conditions/nominal voltage.

On the other hand, increasing the magnetizing current in the winding increases the reactive power of a transformer \citep{price2002geomagnetically, wang2021power}.
As stated above, the DC flux density variation in the core depends on the magnetic reluctance of the DC flux path. 

Therefore, DC flux changes in the cores of three-phase and three-limb transformers are less than in other types because their design creates a higher magnetic reluctance to flow DC in the core. The core material, type of connection, and kind of core have crucial roles in the DC flux effects \citep{mousavi2017analysis,mkhonta2018investigation}.

\section*{Data and Analysis}
\subsection*{Longitudinally symmetric disturbance index}
The symmetric disturbance field in the horizontal direction (SYM-H) has a similar physical unit of \textit{Dst}.  SYM-H is observed by several stations such as San Juan ( 18.11$^{\circ}$ N), Alibag (18.638$^{\circ}$ N), Honolulu (21.32$^{\circ}$ N), Tucson (32.17$^{\circ}$ N), Fredericksburg (38.2$^{\circ}$ N), Boulder (40.13$^{\circ}$ N), Urumqi (43.8$^{\circ}$ N), Memambetsu (43.91$^{\circ}$ N), Chambon-la-Foret (48.025$^{\circ}$ N), Martin de Vivies (37.796$^{\circ}$ S), and Hermanus (34.425$^{\circ}$ S). The SYM-H index provides geomagnetic field disturbance (in nT unit) in mid-latitudes with 1-minute resolution and is accessible at \url{https://wdc.kugi.kyoto-u.ac.jp/} \citep{iyemori2010mid,wawrzaszek2022fractal}.

\subsection*{Magnetograms}
We use the SuperMag ground station measurements to determine the temporal variations of the horizontal geomagnetic field near Iran's geographical coordinates. Magnetogram data include geomagnetic field components in a local NEZ geomagnetic coordinate system (N: north, E: east, Z: down) with a 1-minute resolution. Large fluctuations in the geomagnetic field can signify the emergence of geomagnetic disturbances and GIC in the power grid \citep{rogers2020global}. Equation \ref{mag}  uses magnetometer data to calculate the horizontal geomagnetic field variation ($|dB_{H}/dt|$),
\begin{equation}
\centering
\lvert \frac{dB_{H}}{dt}\rvert=\frac{\sqrt{(\Delta B_{N})^2+(\Delta B_{E})^2}}{\Delta t}\label{mag}    
\end{equation}
where the $\Delta B_{N}$ and $\Delta B_{E}$ are a variation of northward and eastward components of the local geomagnetic field, respectively.
We use SuperMag magnetogram data from IAGA code JAI (26.92$^{\circ}$ N, 75.80$^{\circ}$ E) as a ground station close to Iran territory.

\subsection*{Power grid data}
Power grid data includes the active and reactive powers (the units are in MW and Mvar, respectively) of the transmission transformers taken from Iran Electricity Management Company, which is performed in two voltage classes, 230 kV (522 cases) and 400 kV (197 cases). Iran's power grid data (including active and reactive powers for transformers) is available with one hour resolution from  19 March 2018 to 20 March 2020. To reduce the error, we consider the average active ($P_{A}$) and reactive ($P_{R}$) power of transformers for each substation. The power factor is defined as the ratio of active power to apparent power, $${\rm Power~ factor}=P_{A}/\sqrt{P_{R}^2+P_{A}^2}.$$
Some power grid equipment, such as capacitors or SVC, etc., compensate for the power grid's different parts for a power factor between 0.9 and 0.95. A decrease in the power factor of around 0.85 indicates the non-optimal use of the capacity of the grid's equipment and the lack of enough compensation.

\begin{figure}[h]
    \centering
    \begin{subfigure}{.6\textwidth}   
        \hspace*{-11.5cm}
        \vspace*{-11.cm}
        \includegraphics[width=3.3\textwidth]{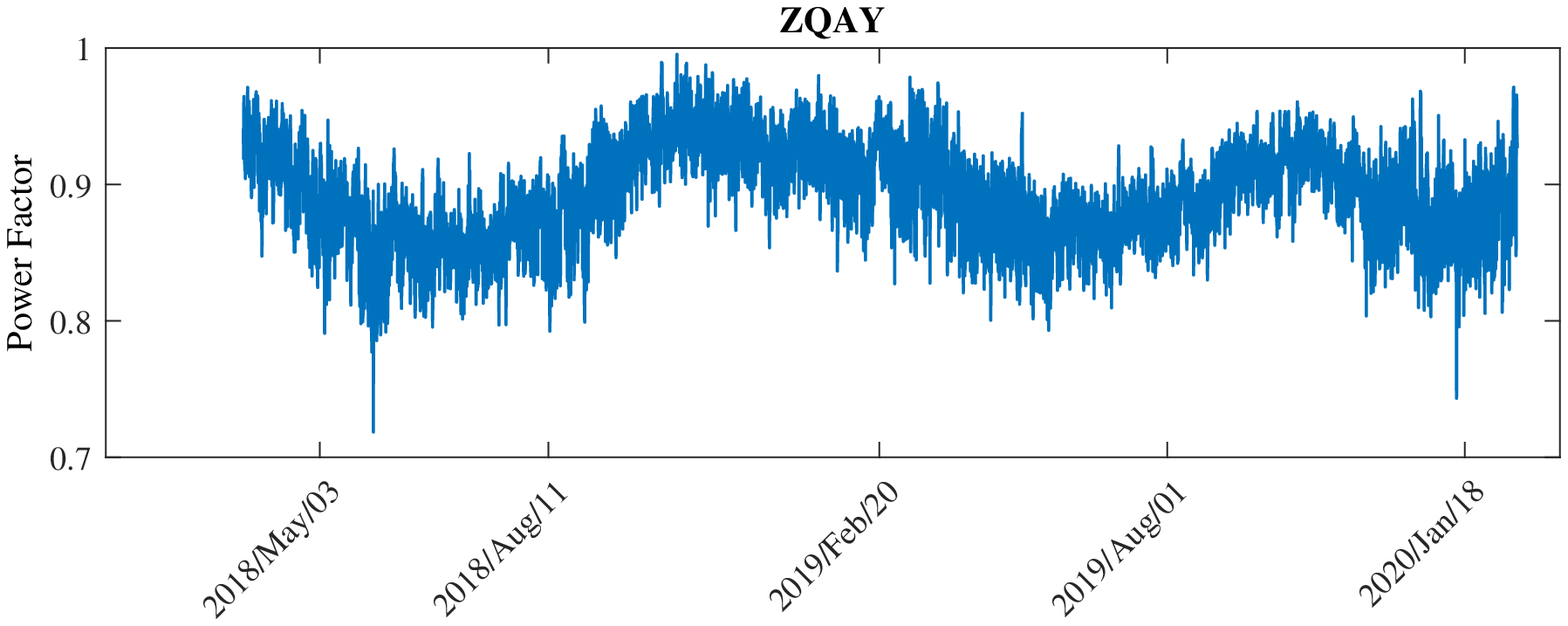}
       \end{subfigure}
    \begin{subfigure}{.6\textwidth}
    \hspace*{-11.5cm}
    \vspace*{1.cm}
        \includegraphics[width=3.3\textwidth]{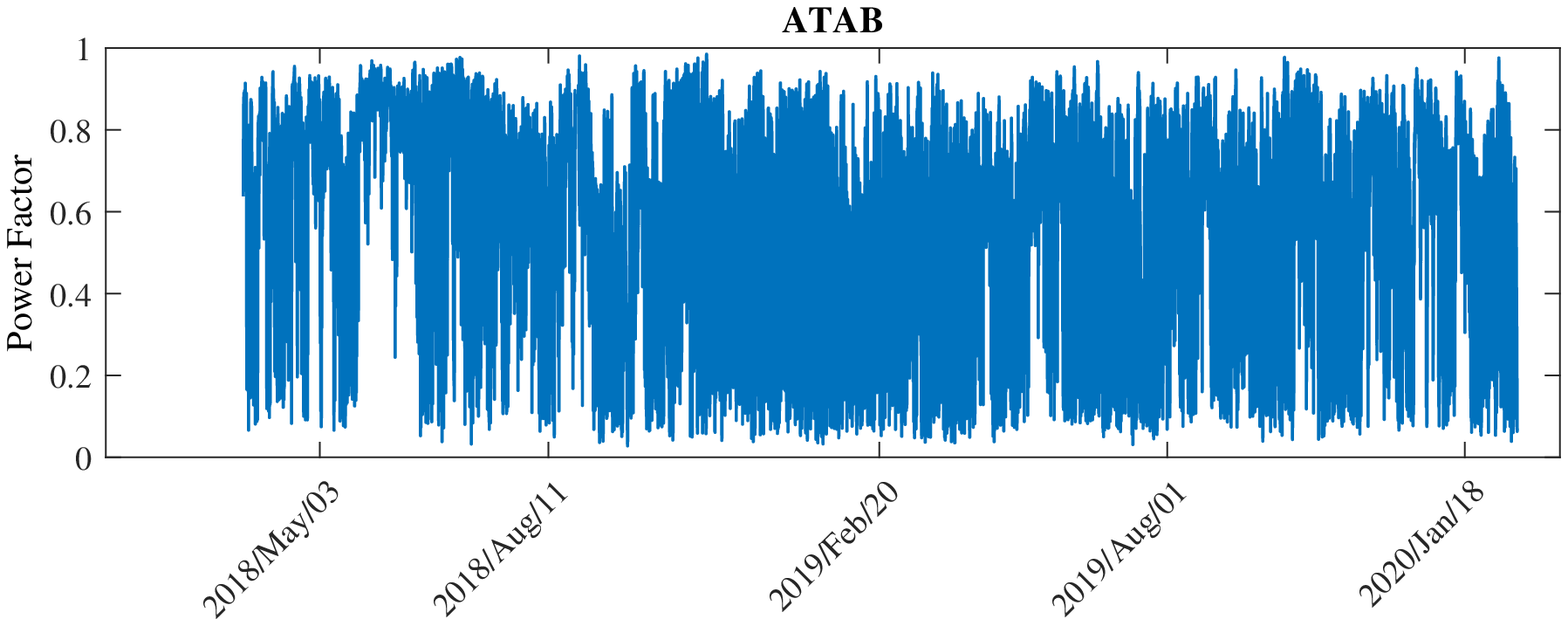}
    \end{subfigure}
    \vspace*{-12.cm}
    \caption{Samples of time series of power factor of the 400 kV transformers (top) with power factors mostly varies in the range of  0.7 and 1, (bottom) with power factors mostly varies in the range of 0.0 and 0.7.}
    \label{fig3}
\end{figure}

Figure {\ref{fig3}} shows the time series of power factors for two 400 kV transformers. In the first case (Figure {\ref{fig3}} top), the power factor is in the range of 0.7 to 1 for a transformer as a part of a network without the neutral points. But, in the second case (Figure {\ref{fig3}} bottom), we indicate the power factor for a transformer with neutral points in the network, which shows the power factor is mostly below 0.7. So, we exclude the power factor for transformers with many variations below 0.7. Hereafter, we analyze the power factor of transformers is similar to the first case, in that most of the power factors vary above 0.7.  

Reducing the power factor to less than 0.7 or increasing the reactive power compared to a normal situation depends on several factors in the network structure or external factors. In synchronizing it with a geomagnetic storm, the GIC may be one of the effective candidates for the power values perturbation. The GIC enters the power grid from the neutral point of the transformers and flows in the transmission lines. To this end, we investigate the correlations of the SYM-H index reduction to less than -30 nT and the lack of optimal power distribution (power factor less than 0.7) in the electric power network.

\begin{figure}[h]
    \centering
    \includegraphics[width=1.\textwidth]{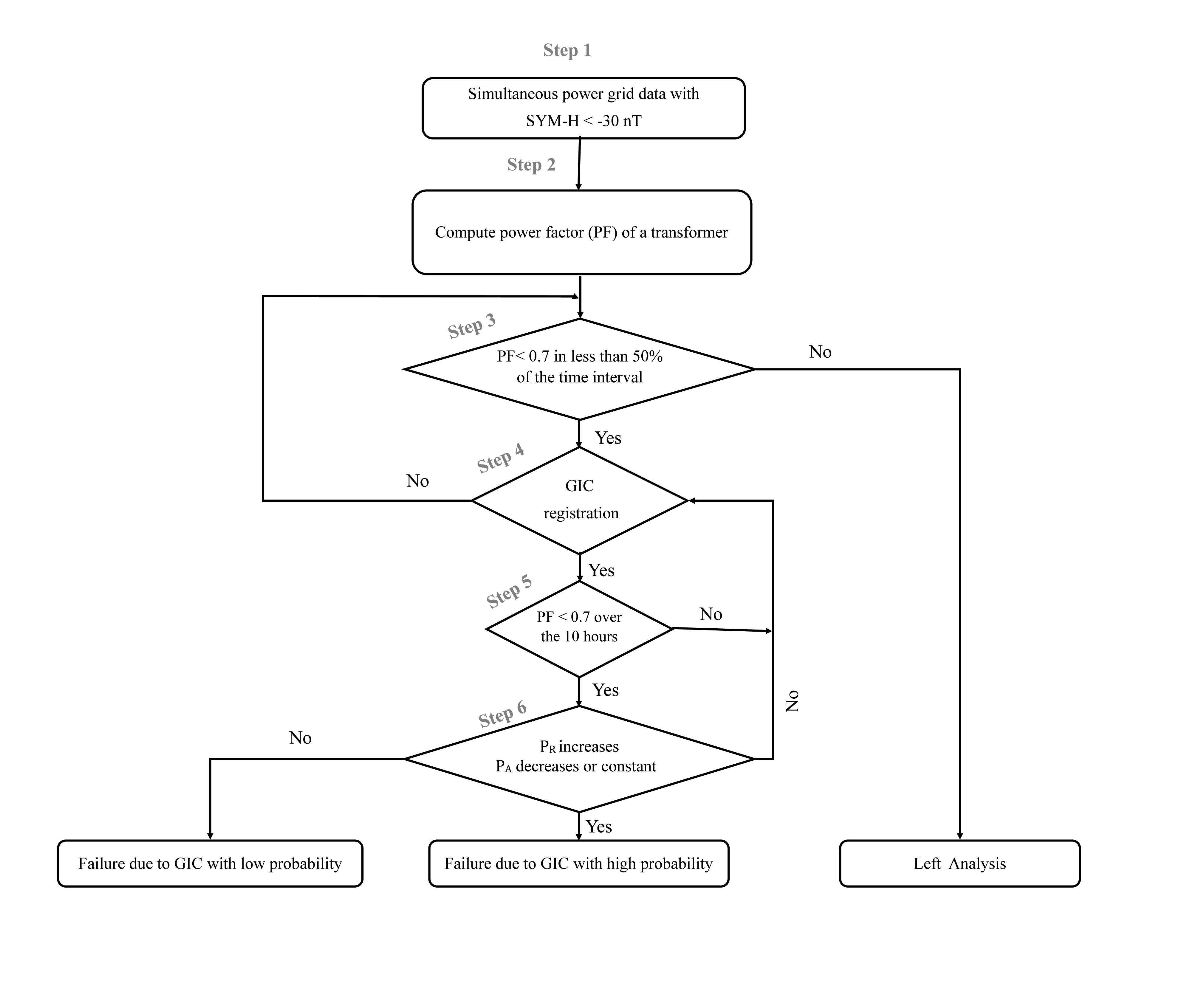}
    \caption{An algorithm to identify the power grid transformers affected by GIC.}
    \label{fig4}
\end{figure}

Figure \ref{fig4} illustrates the algorithm for estimating the GIC in the power network. First, for the power grid data correlated with the SYM-H $<$ -30 nT, we calculate the power factor for two classes of 230 and 400 kV transformers (Steps 1 and 2). If the value of the power factor is less than 0.7 in 50 percent of the given time, we consider this transformer as one of the potential transformers affected by GIC (Steps 3 and 4). In the next step, if the power factor reduction to less than 0.7 continues for 10 hours, the transformer remains on the list of affected ones (Step 5). However, it is necessary to check the amount of active and reactive power during this period. If the reactive power increases compared to the normal state and the active power decreases or remain constant, the GICs in this transformer can be confirmed with a high probability (Step 6).

\section*{Results}
For March 2018 to March 2020, we detect 128,627  cases (with a power factor less than 0.7) for Iran's power network, of which 12,112 correlate with GMD for SYM-H  $<$ -30 nT (flowchart of Figure \ref{fig4}). \citet{gil2021evaluating} showed that out of 25,616 and 30,155 registered failures of the electrical grids in southern Poland, 4625 and 10,656 failures might relate to GMD in 2010 and 2014, respectively. The total number of failures near the solar maximum (January–July 2014) is twice that of the solar minimum (2010). Therefore, their results confirm the positive correlation between the number of electrical grid failures and solar activity.

\begin{figure}[h]
    \centering
\hspace*{-1.cm}
    \includegraphics[width=0.8\textwidth]{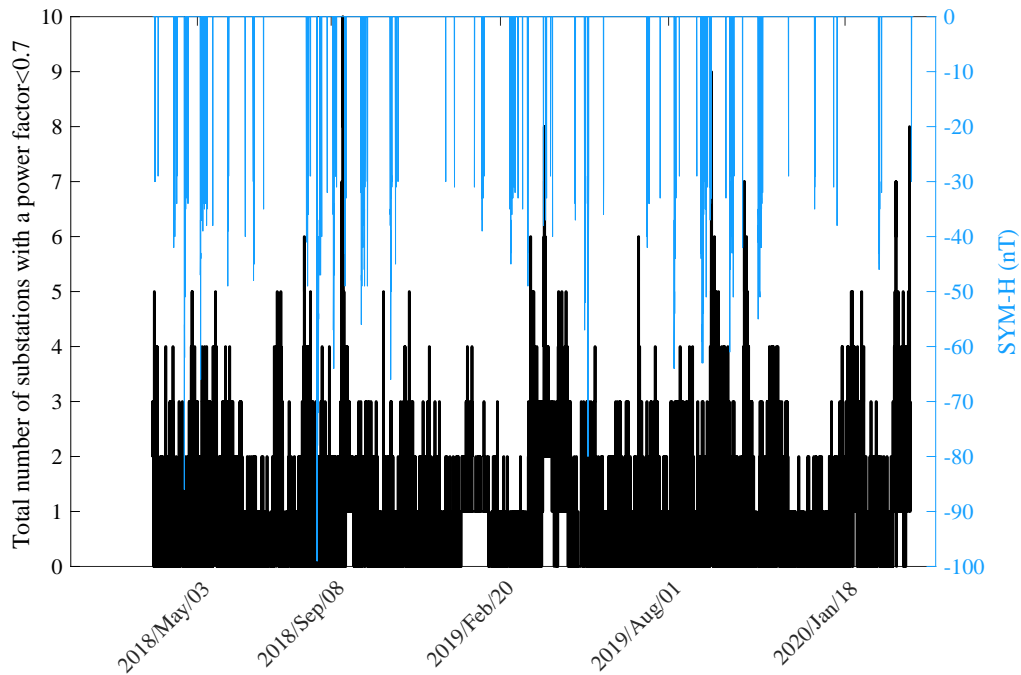}
    \caption{Total number of substations with power factor below 0.7 at every time along with SYM-H $<$ -30 nT.}
    \label{fig5}
\end{figure}

Figure {\ref{fig5}} shows the number of Iran substations (with a power factor less than 0.7) and the time series of the SYM-H $<$ -30 nT at every time. We eliminate the statistics of the turn-off substations with zero active and reactive powers. The power factor and SYM-H cadences (resolutions) are 1 hour and 1 minute, respectively. We find that only at 4 percent of times within two years, the SYM-H has values less than -30 nT. To determine the possible impact of geomagnetic activity on the power network, we consider the time interval of ten hours (flowchart of Figure \ref{fig4}) after each SYM-H with value $<-30$ nT. We select transformers with a power factor less than 0.7 at less than 5 percent of analyzing period. In other words, these transformers deliver high performance (power factor $>$ 0.7) at 95 percent of working times. We observe that these transformers with a power factor less than 0.7 show at least a 55 percent correlation (moderate) with SYM-H $<$-30 nT. This finding may indicate the GIC in the power network due to disturbance of the Earth's magnetic field, which needs to confirm by installing the appropriate instrument in Iran's power grid substations. \citet{zhang2020measurements} applied the SYM-H index in analyzing the GICs measurement at China's low‐latitude power grid substations and found a positive correlation between the large GIC and considerable change in the horizontal magnetic field.

\begin{figure}[h]
    \centering
    \hspace*{-1.cm}
\vspace*{-1.cm}
   \includegraphics[width=0.8\textwidth]{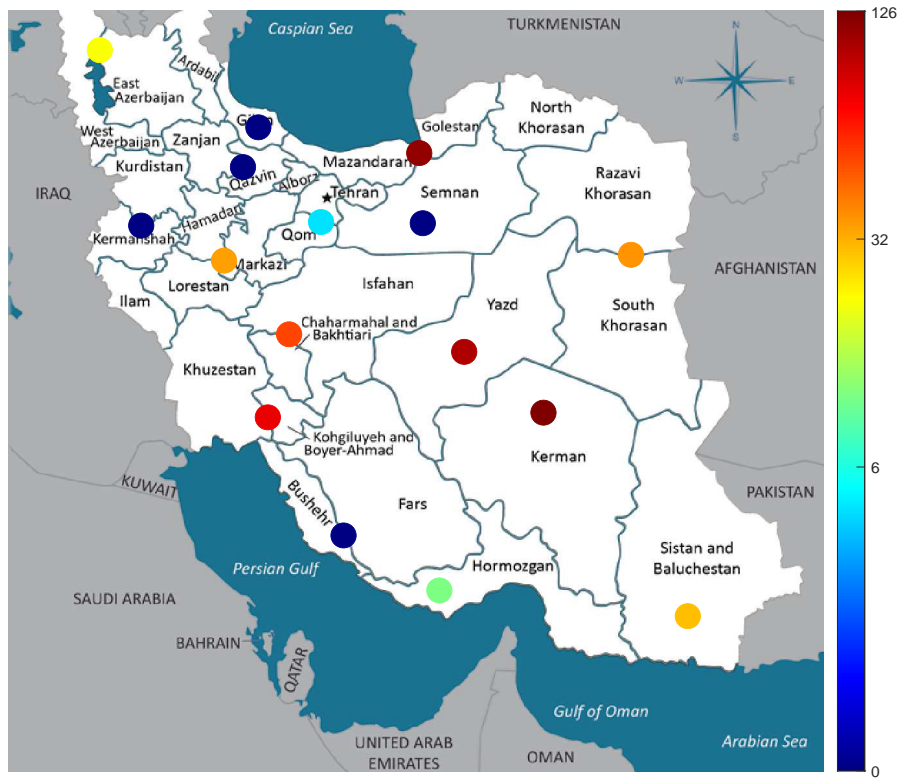}
    \caption{Color map of the possible impact of geomagnetic storms on Irans' 230 and 400 kV transformers. The color bar values indicate the ratio of the number of power factors ($<$ 0.7) correlated (flowchart of Figure \ref{fig4}) with SYM-H ($<$ -30 nT) to the number of transforms in the region within two years.}
    \label{fig6}
\end{figure}

 Following the algorithm in Figure \ref{fig4}, we extracted the power factor of less than 0.7 for two classes of 230 and 400 kV transformers. Then we considered their correlations with SYM-H $<$ -30 nT (time interval less than ten hours). Figure {\ref{fig6}} represents the color map of the possibility of the lack of optimal power correlated with SYM-H $<$ -30 nT within 2018-2020 and in different regions of Iran's select regions with a blue circle have the lowest possibility of being at risk from geomagnetic storms. 

\begin{figure}[h]
    \centering
    \hspace*{-3.cm}
    \includegraphics[width=0.6\textwidth]{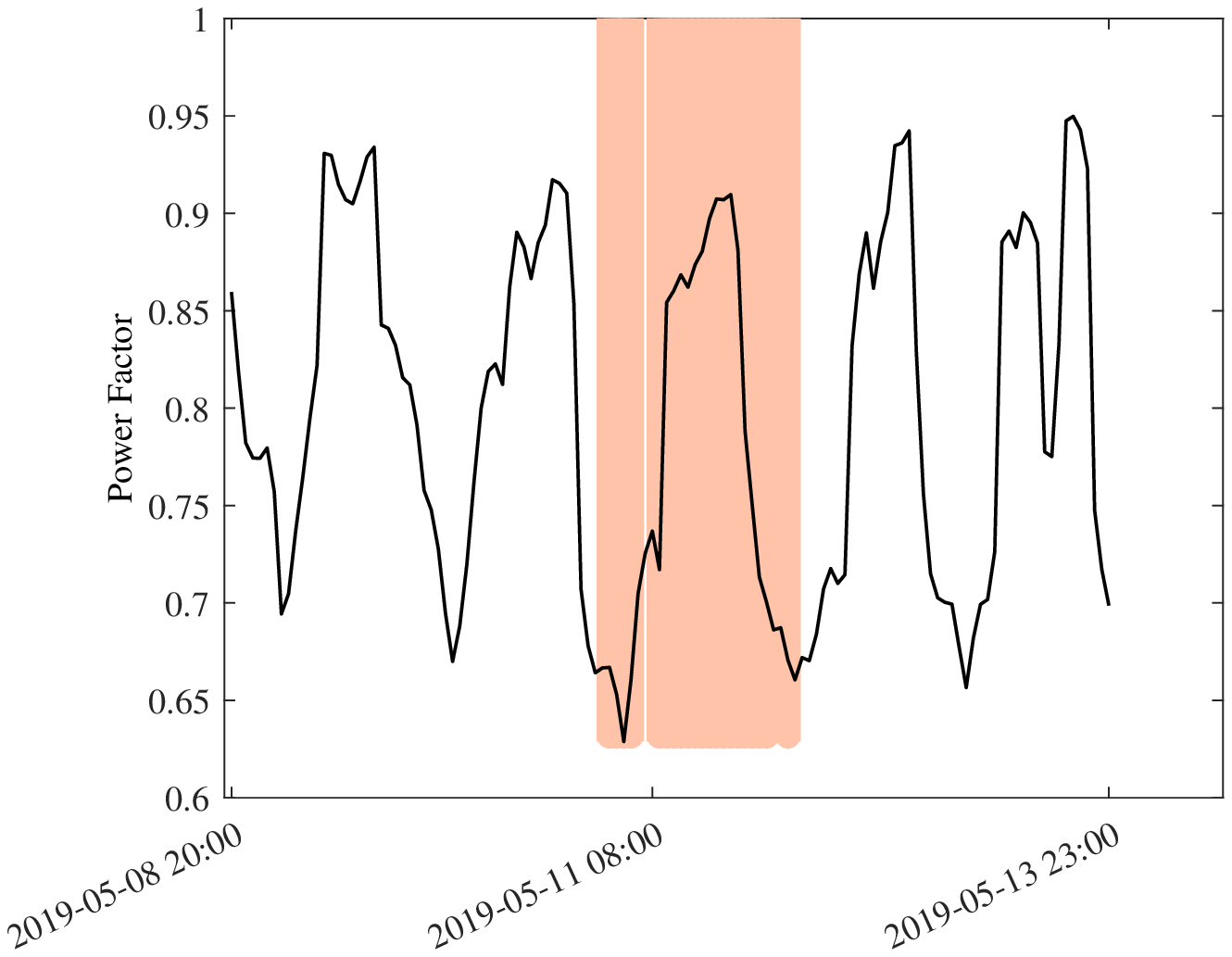}
   \caption{Oscillating perturbations of power factor in a 230 kV transformer. Pinked strip shows time interval SYM-H$<$-30 nT.}
    \label{fig7}
\end{figure}

For more accurate analysis, we eliminate the time interval when oscillating changes in power factor accompany the lack of optimal power distribution in the network. Figure {\ref{fig7}} shows a sample of oscillating power factor changes for a 230 kV transformer. By visual inspection, we select non-oscillatory perturbations (power factor less than 0.7) correlated with SYM-H $<$ -30 nT. Our investigations show that 89 and 84 cases of disturbances may originate from geomagnetic storms in 400 kV and 230 kV transformers, respectively.

\begin{figure}[h]
    \centering
    \hspace*{3.cm}
    \includegraphics[width=1.1\textwidth]{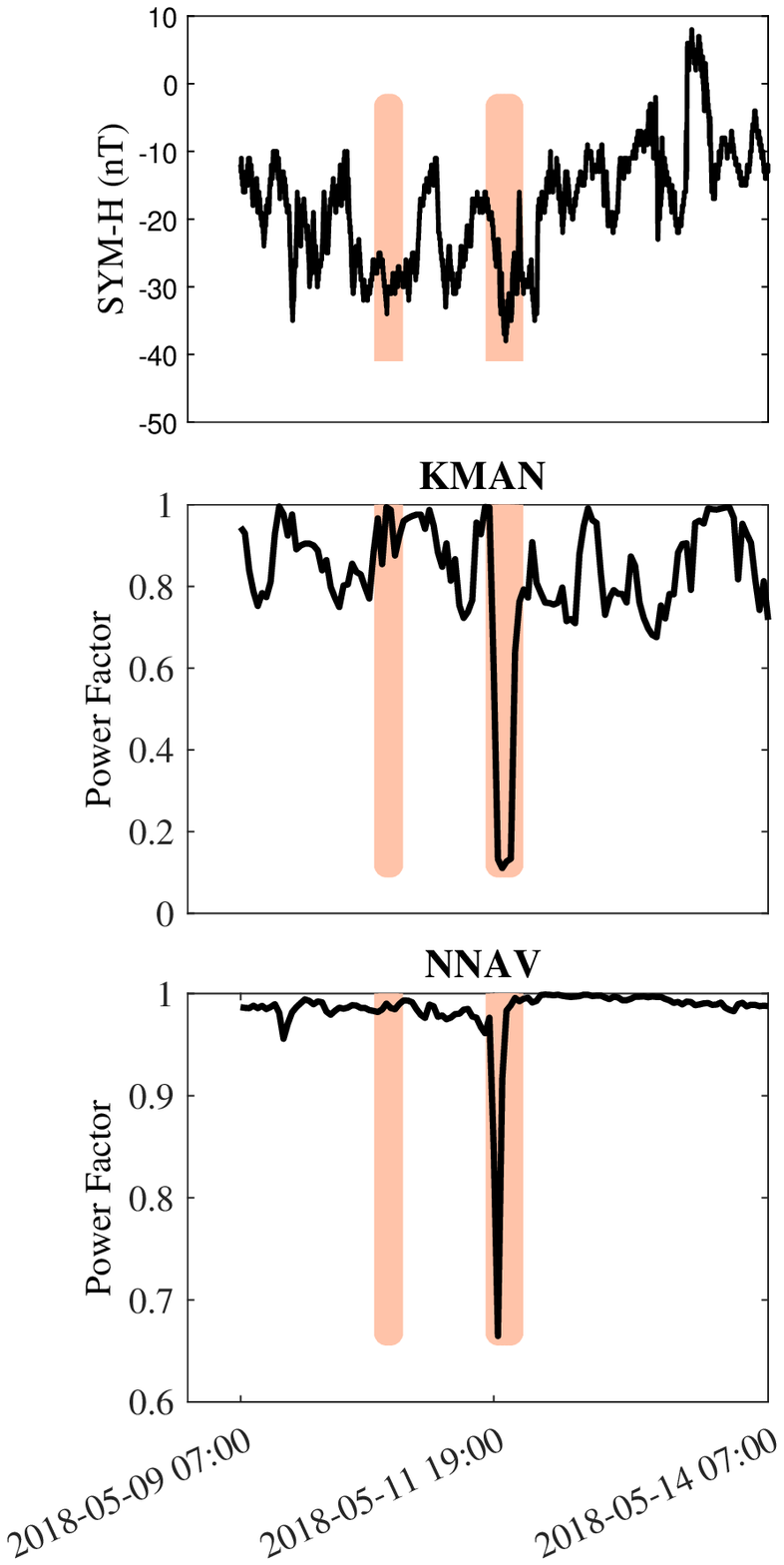}
    \caption{Time series of SYM-H index (upper panel) and power factor related to posts KMAN (middle panel) and NNAV (bottom panel). Pinked strips show the time of occurrence of the geomagnetic storm.}
    \label{fig8}
\end{figure}

Synchronization and connection between the transformers' perturbations can be two important characteristics of increasing their correlation with geomagnetic storms. Forty-one times, these perturbations appear in at least two transformers. Figure {\ref{fig8}} presents the time series of the SYM-H index (upper panel) and the power factor of Manoujan (KMAN) (middle panel) and Navard (NNAV) (bottom panel) transformers' electrical substations. As shown in the figure, the power factor of the transformers in two substations decreased to less than 0.7 at the same time (11 May 2018).

\begin{figure}[h]
    \centering
    \hspace*{3.cm}
    \includegraphics[width=1.1\textwidth]{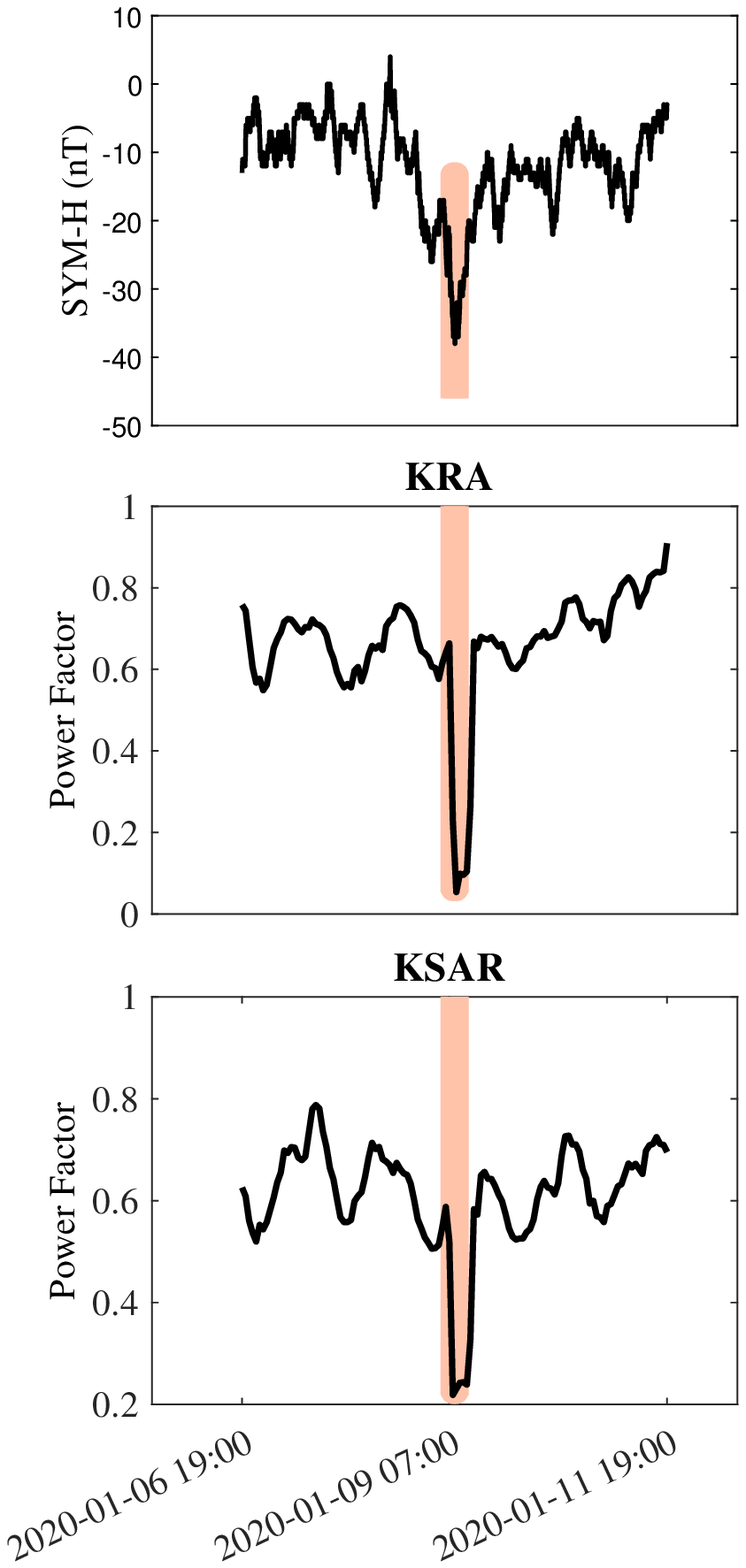}
    \caption{Time series of SYM-H index (upper panel) and power factor related to substation KRA (middle panel) and KSAR (bottom panel). Pinked strips show the time of occurrence of the geomagnetic storm.}
    \label{fig9}
\end{figure}

\begin{figure}[h]
    \centering
   \includegraphics[width=0.8\textwidth]{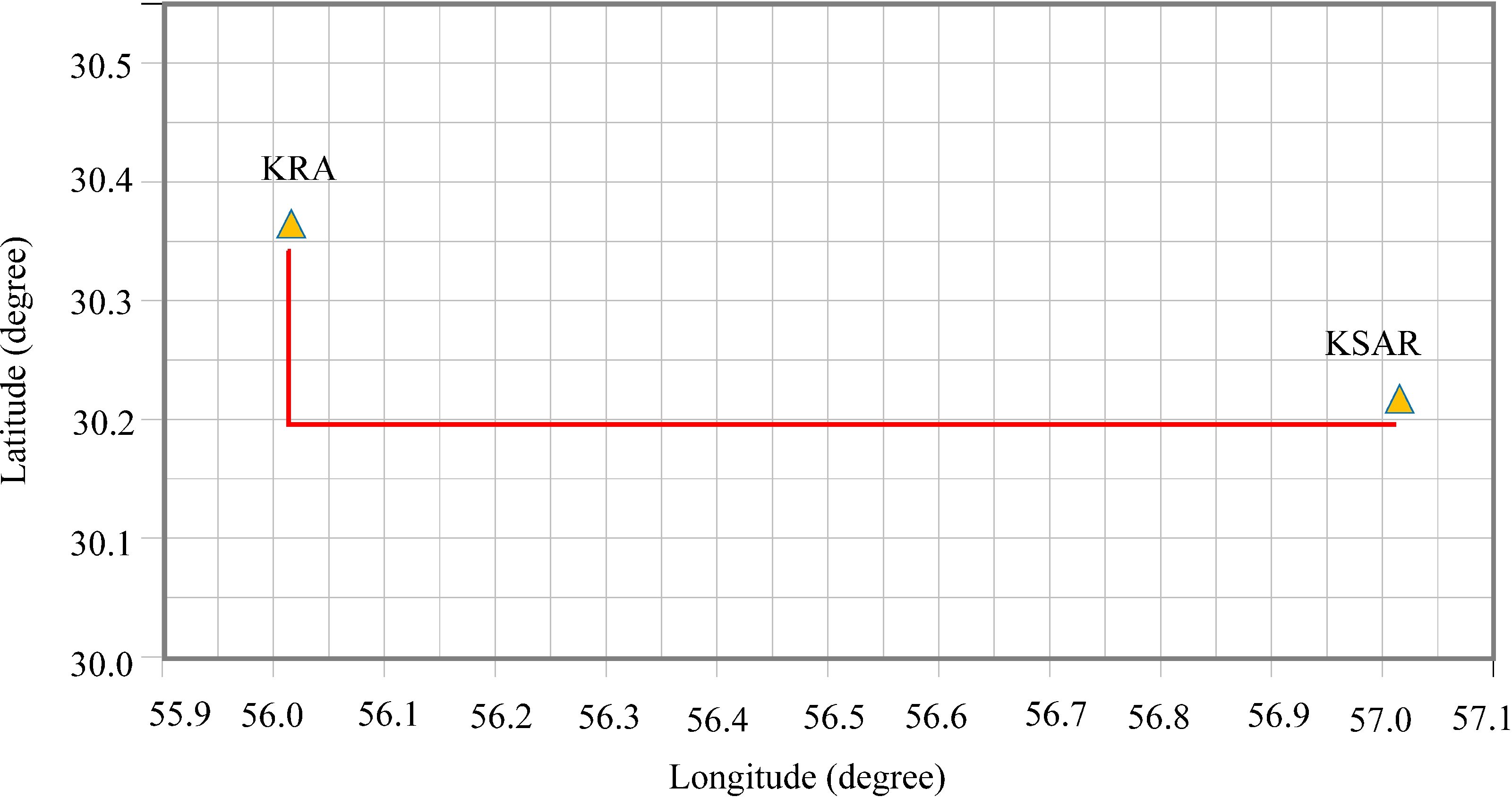}
    \caption{Geographic locations of the KRA and the KSAR substations (Orange triangles) and the transmission line (red line) in the north of Iran.}
    \label{fig10}
\end{figure}

The connection of two substations and a decrease in the substations' power factor during the geomagnetic storm are essential factors to verify the GIC in the network. These two factors may confirm the correlation with the storm. Figure {\ref{fig9}} presents the time series of the SYM-H index (upper panel) and transformers power factor of Rafsanjan (KRA) (middle panel) and Sarcheshmeh (KSAR) (bottom panel) substations with common transmission lines. Figure \ref{fig10} presents the geographical location of the Rafsanjan-Sarcheshmeh transmission line. The transmission line length of the two stations is 56 km, located in the east-west direction. The end of the connecting lines of two substation transformers are wye configuration and have a neutral point for GIC flow.

\begin{figure}[h]
    \centering
    \hspace*{3.cm}
    \includegraphics[width=1.2\textwidth]{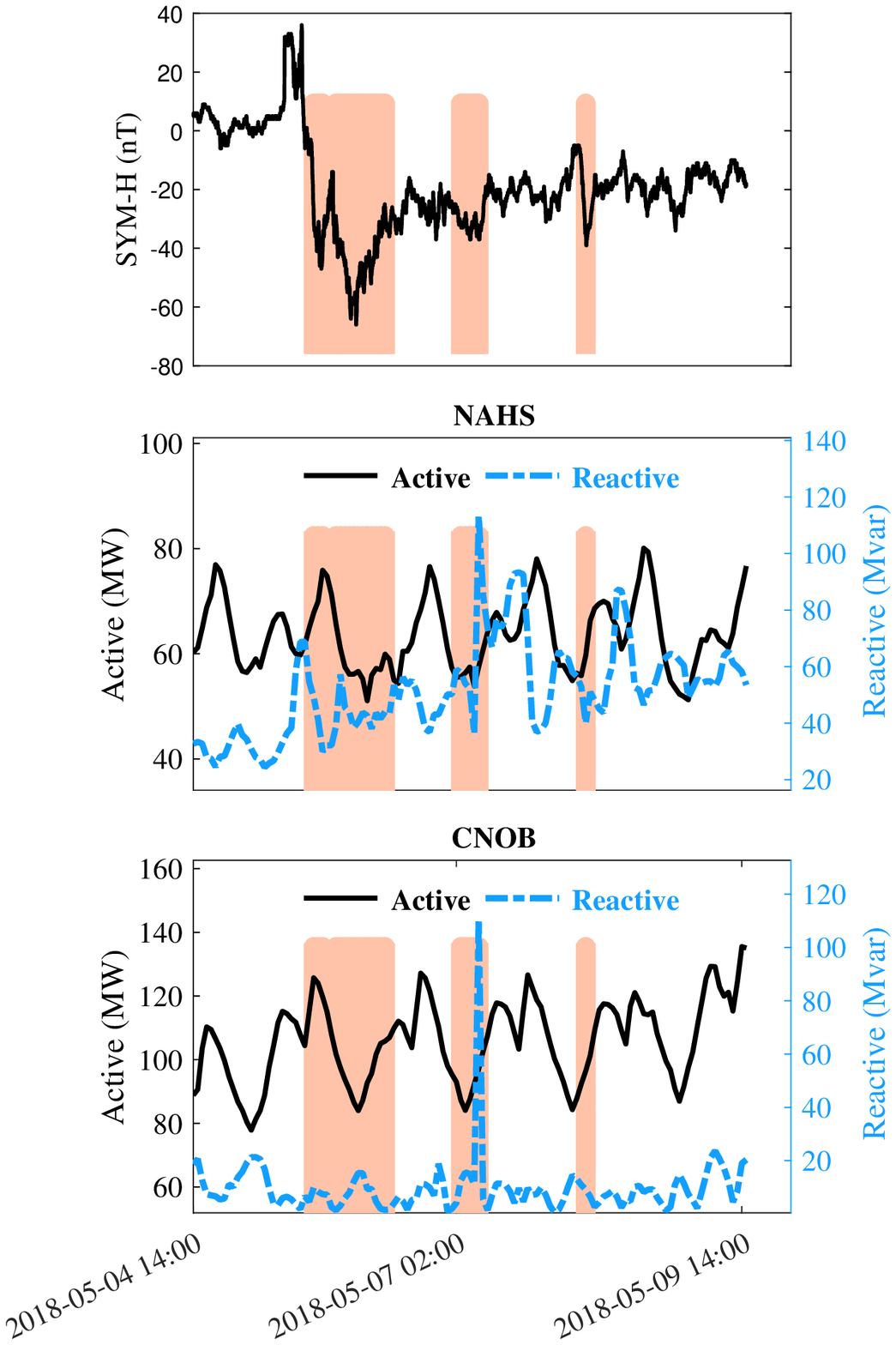}
    \caption{Time series of SYM-H index (upper panel) and power factor, active and reactive power related to substations NAHS (middle panel) and CNOB (bottom panel). Pinked strips show the time of occurrence of the geomagnetic storm.}
    \label{fig11}
\end{figure}

Decreases in the power factor to less than 0.7 is due to increasing the reactive power and reducing (or remaining constant) the active power to lower values.  A considerable decreasing power factor may be considered as evidence for GICs in the transformers. By visual inspection, we find the reactive power peaks above optimal values correlated with SYM-H $<$ -30 nT. Our investigation shows that a geomagnetic storm may cause 40 and 44 cases in 400 kV and 230 kV transformers, respectively. In 21 cases, these highly appear in at least two transformers. Figure {\ref{fig11}} presents the time series of SYM-H index (upper panel) moreover, active and reactive powers related to NAHS (middle panel) and  CNOB (bottom panel) transformers. As shown in the figure, active power remains constant, but the reactive power is significantly increased and correlated with SYM-H (7 May 2018).

By analyzing the solar event catalog (\url{https://www.sidc.be/cactus/catalog.php}), we find a set of coronal mass ejections (CMEs) appeared on 3 May 2018 and 9 May 2018, which may have caused the GMDs. Typically, due to the velocity and angular width of a CME, it may reach the Earth's environment within a few days and then affect the Earth's magnetic field. Quantitative analysis of solar energetic particles detected by near the Earth instruments (GOES, ACE, etc.) may identify their solar event sources \citep[e.g.,][]{Mohammadi2021JGRA}.

\citet{wang2021simulation} suggested a simulation method to recognize the critical substations of the China power grid affected by GIC. Their study provides monitoring to control the damage to the China power network. A similar simulation of Iran's power grid structure needs to be employed  for identifying the critical substations that may be affected by GIC.    

\begin{figure}[h]
    \centering
    \hspace*{3.cm}
    \includegraphics[width=1.2\textwidth]{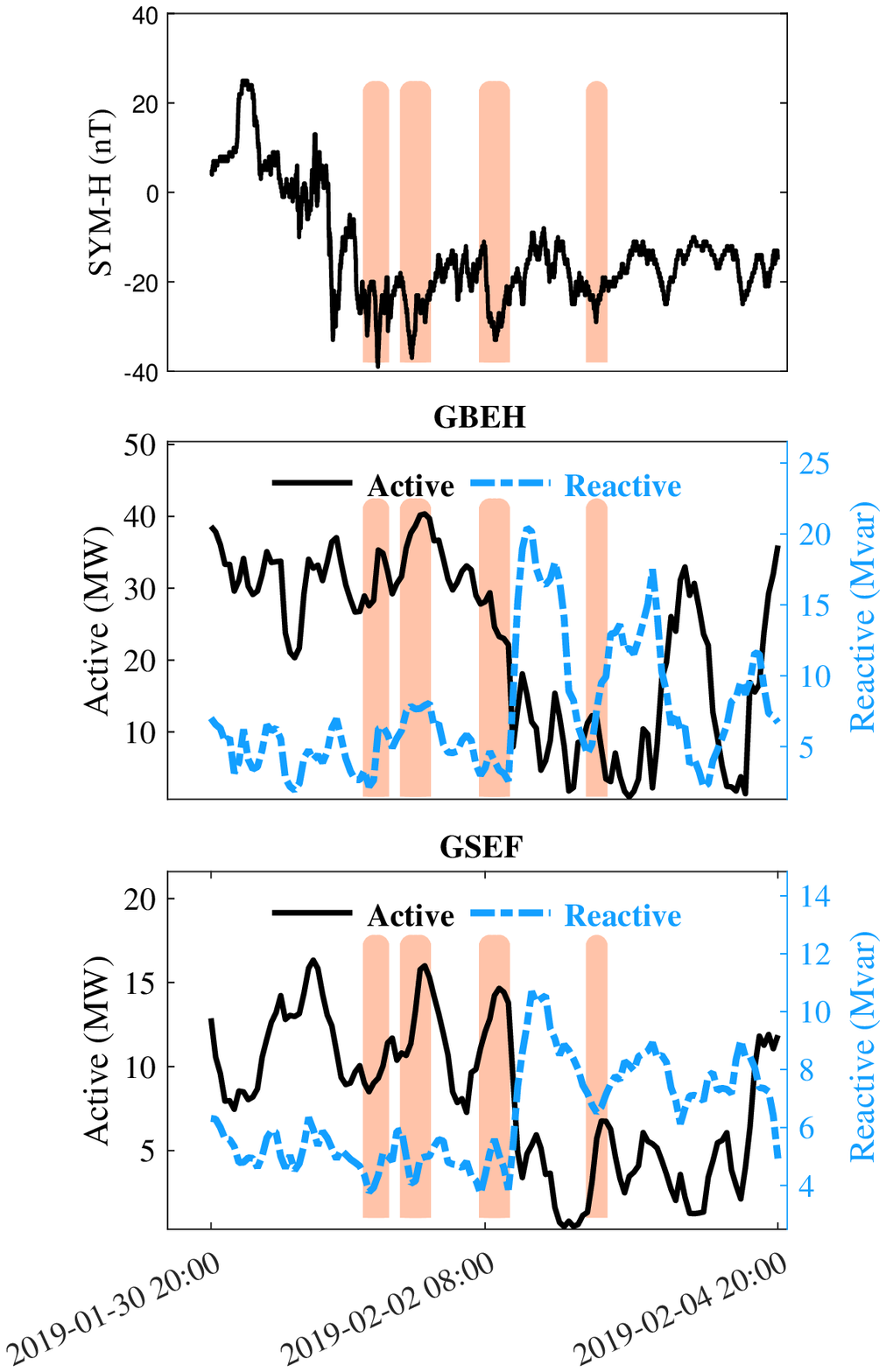}
    \caption{Time series of SYM-H index (upper panel) and active and reactive power related to substations GEH (middle panel) and GSEF (bottom panel). Pinked strips show the time of occurrence of the geomagnetic storm.}
   \label{fig12}
\end{figure}

\begin{figure}[h]
    \centering
\includegraphics[width=0.8\textwidth]{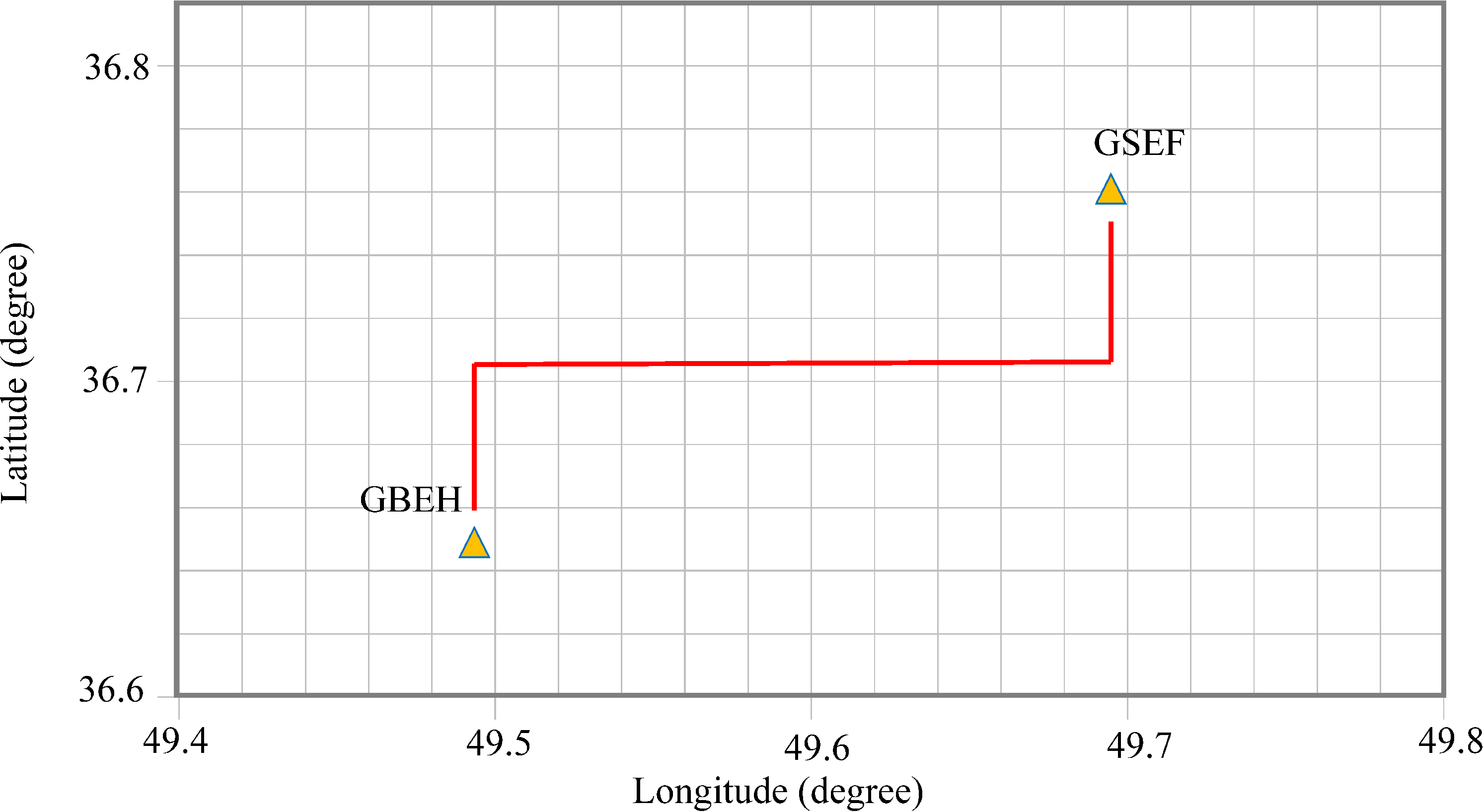}
    \caption{Geographic locations of the GSEF and the GBEH substations (Orange triangles) and the transmission line (red line) in the north of Iran.}
    \label{fig13}
\end{figure}

The connection of two substations, a decrease in substations' active power, and an increase in reactive power during the geomagnetic storm are essential factors to verify the GIC in the network. These factors may confirm the correlation with the storm. Figure {\ref{fig12}} presents the time series of the SYM-H index (upper panel) and transformers power factor of Sefidrood (GSEF) (middle panel) and Shahid-Beheshti (GBEH) (bottom panel) substations with common transmission lines. Figure \ref{fig13} presents the geographical location of the Sefidrood-Shahid-Beheshti transmission line. The transmission line length of the two stations is 14 km, located in the east-west direction. The end of the connecting lines of two substation transformers are wye configuration and have a neutral point for GIC flow.

\begin{figure}[h]
\hspace*{1.cm}
\vspace*{-2.cm}
    \centering
\includegraphics[width=1.5\textwidth]{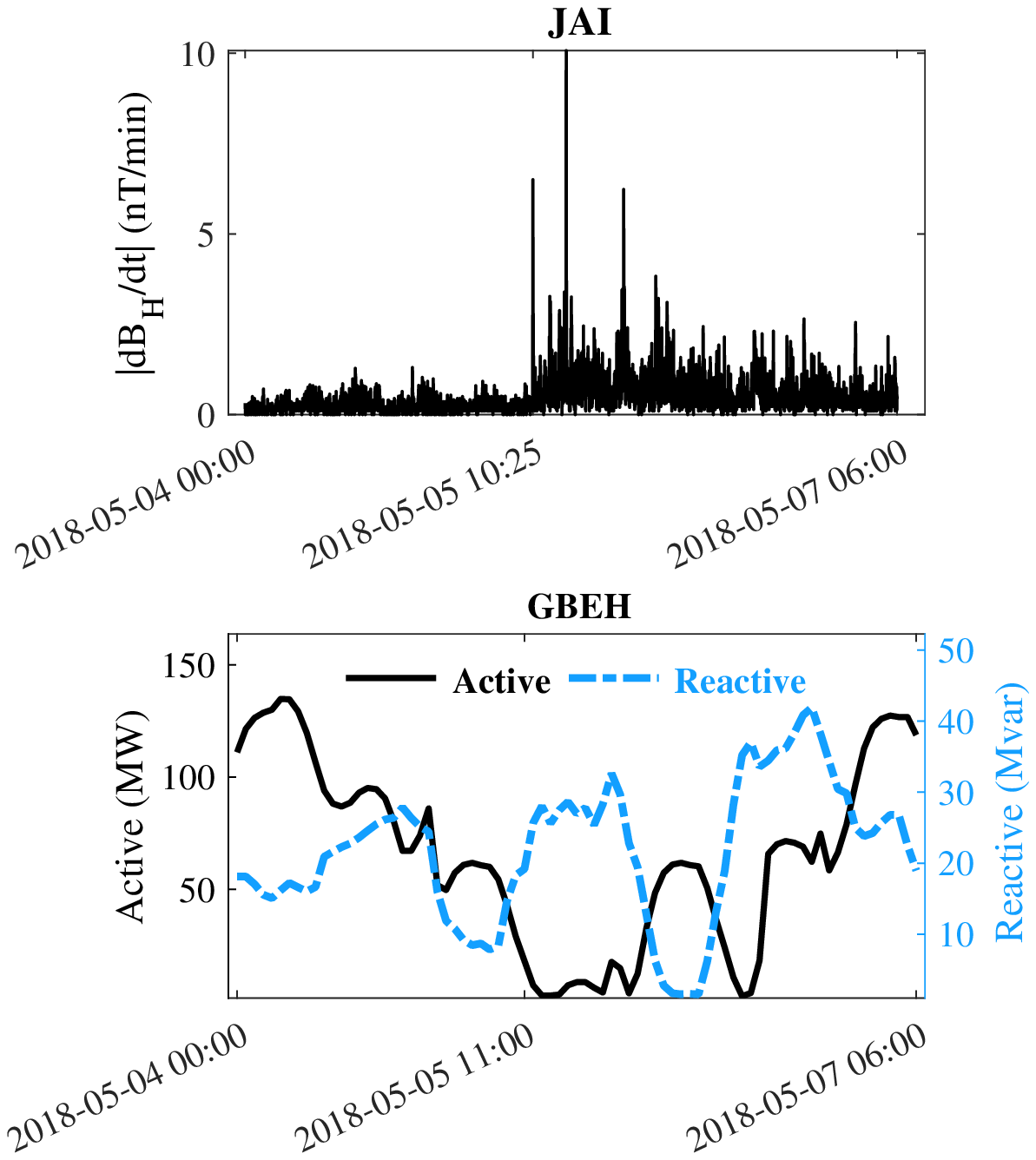}
    \caption{Time series of horizontal magnetic field variation $|dB_H/dt|$ recorded by JAI station (upper panel), active and reactive powers for substation GBEH (bottom panel) from 4 May 2018 to 7 May 2018.}
    \label{fig14}
\end{figure}

The horizontal geomagnetic field variation recorded by the local magnetic observatories is an essential indication of a magnetic storm. The lack of such information was due to the lack of well organized local magnetic field observatories in Iran, so we used the data from JAI as one of the nearest observatories. Figure \ref{fig14} shows a sample of time series of horizontal geomagnetic field variation observed by JAI (26.92 $^\circ$ N, 75.8$^\circ$ E) and transformers power factor of Shahid-Beheshti (GBEH) substation. A sizeable geomagnetic field fluctuation correlates with increasing reactive power and decreasing active power in Shahid-Beheshti (GBEH) substation. The horizontal geomagnetic field variations may affect the power network and probably lead to the GIC \citep{dods2015network,rogers2020global}.

We observe that some reactive power correlated with SYM-H, but others did not respond to SYM-H. This inconsistency may be due to the characteristics of SYM-H as a global index for GMDs. Therefore, a complete correlation between reactive powers and SYM-H could not be expected. Some reactive powers are well correlated with the local magnetic field measurements, which shows the necessity of ground base observatories to record the local magnetic field variation.

\section{Conclusions}

GICs flowing in the power systems is one of the GMD effects as a space weather indicator which can cause saturation of power transformers. The SYM-H index is one of the most appropriate parameters to investigate the effect of geomagnetic storms. In this article, by analyzing the powers of transformers, we examined the evidence of GIC flux in the network. In data processing, according to the algorithm in Figure \ref{fig4}, we considered each substation as a node in the network. We used the average power value of all transformers in a substation for data analysis.

Our investigation covering the period from 19 March 2018 to 20 March 2020 shows that Iran's power network, a mid-latitude system, experienced a total of 128,627 cases (with a power factor less than 0.7), out of which 12,112 samples were likely due to GMDs. These results may imply the presence of GIC in the Iran power network due to GMDs. Our analysis for high-performance transformers (with a power factor greater than 0.7) showed that the lack of optimal distribution in 55 percent correlates well with SYM-H values less than -30 nT.

In the local magnetic field data analysis, we found that considerable changes in the horizontal field significantly correlate with an increase in the reactive power at the GBEH substation. Although there are no proper installations to measure the GICs in Iran's power grid, two sample substations show probable evidence of GIC in the network due to geomagnetic storms. The first event appeared from the Rafsanjan to Sarcheshmeh substations (with wye configurations) with a 56 km transmission line length correlated with a moderate geomagnetic storm (SYM-H=-31 nT). The second event from the Sefidrood to Shahid-Beheshti substations (with wye configurations) and 14 km transmission line length coincided with a geomagnetic storm of moderate intensity (SYM-H=-34 nT).

We studied the active power, reactive power, and power factor of Iran's power network within two years of the minimum solar activities, which indicates the effect of geomagnetic storms in the network. For proper investigation of the effects of GMDs on Iran's power grid, we require long-term studies and some more transformer parameters such as voltage, current, and temperature with tiny time intervals. The long-term studies were done in Brazil and Japan as the middle latitude countries affected by a geomagnetic storm \citep{trivedi2007geomagnetically,watari2015estimation}.

Many studies \citep[e.g.,][]{kirkham2011geomagnetic,liu2013preliminary,sokolova2018algorithm,wang2019machine,zhang2020mitigation,albert2022analysis}  show that countries that have developed the space weather monitoring infrastructure to predict and recognize GMD events may prevent damage to their electric-based instruments. Developing solar and space weather monitoring centers helps to predict and protect against the effects of geomagnetic disturbances in middle-east countries that are on the road to development and space investigations. 

\clearpage
\newpage
\section{Acknowledgments}
The authors thank SuperMAG collaborators (\url{https://supermag.jhuapl.edu/info/?page=acknowledgement}, last access: 22 June 2021) for magnetometer data. Data of SYM index (\url{http://wdc.kugi.kyoto -u.ac.jp}) is acknowledged.

\bibliographystyle{jasr-model5-names}
\biboptions{authoryear}
\bibliography{bibo}     

\end{document}